%% file: TRDM.tex
\documentclass[12pt]{article}
\usepackage{amssymb}

\usepackage{amsmath}
\usepackage{booktabs}
\usepackage{graphicx}
\usepackage{theorem}

\newtheorem{thm}{Theorem}[section]

\newtheorem{prp}{Proposition}[section]
\theorembodyfont{\rmfamily}

\makeatletter

\@addtoreset{equation}{section}
\makeatother

\input{anot.tex}

\begin{document}

\title{Small Area Predictors with Dual Shrinkage of Means and Variances}

\author{
Hiromasa Tamae\thanks{Graduate School of Economics, University of Tokyo, 7-3-1 Hongo, Bunkyo-ku, Tokyo 113-0033, JAPAN, E-Mail: taisho.1603@gmail.com}
and
Tatsuya Kubokawa\footnote{Faculty of Economics, University of Tokyo, 7-3-1 Hongo, Bunkyo-ku, Tokyo 113-0033, JAPAN. \newline{ E-Mail: tatsuya@e.u-tokyo.ac.jp}}
}
\maketitle
\begin{abstract}
The paper  concerns small-area estimation in the Fay-Herriot type area-level model with random dispersions, which models the case that the sampling errors change from area to area.
The resulting Bayes estimator shrinks both means and variances, but needs numerical computation to provide the estimates.
In this paper, an approximated empirical Bayes (AEB) estimator with a closed form is suggested.
The model parameters are estimated via the moment method, and the mean squared error of the AEB is estimated via the single parametric bootstrap method.
The benchmarked estimator and a second-order unbiased estimator of the mean squared error are also derived.
\par\vspace{4mm}
\end{abstract}

\section{Introduction}

Small area estimation (SAE) using linear mixed models has been extensively studied in the literature from both theoretical and applied points of view.
For a good review and account on this topic, see Ghosh and Rao (1994), Pfeffermann (2002), Rao (2003) and Datta (2009).
Of these, the Fay-Herriot model introduced by Fay and Herriot (1979) has been used as an area-level model in SAE.

\medskip
Suppose that there are $m$ small areas and that $y_1, \ldots, y_m$ are direct estimates of small area means.
The Fay-Herriot model is described as 
\begin{align*}
y_i \mid \xi_i \sim& \Nc(\xi_i, \si_i^2),
\\
\xi_i \sim& \Nc(\z_i^T\bbe, \tau^2),
\end{align*}
where $\z_i$ is a vector of auxiliary variables and $\bbe$ is an unknown vector of regression coefficients.
Although $\si_i^2$'s are treated as known variances in the Fay-Herriot model, in practice, $\si_i^2$ are estimated quantities, and the resulting empirical Bayes (EB) estimators involve substantial estimation errors.
To take this point into account, we suppose that statistics $V_1, \ldots, V_m$ are available for estimating $\si_i^2$ and that $V_i/\si_i^2$ has a chi-square distribution with $n_i$ degrees of freedom.
Then, Wang and Fuller (2003) provided estimators of the mean squared error (MSE) of the empirical Bayes estimators.
For such variance modeling approaches, see Arora and Lahiri (1997), You and Chapman (2006), Dass, Maiti, Ren and Sinha (2012), Jiang and Nguyen (2012).
Also see Maiti, Ren and Sinha (2014) and the references therein.

\medskip
In the Fay-Herriot models with heteroscedastic unknown variances, each variance $\si_i^2$ cannot be estimated consistently based on $V_i$ when $n_i$'s are bounded.
This leads to the inconsistency properties of estimation procidures, namely, the empirical Bayes estimator does not converge to the Bayes etimator, and the MSE of the empirical Bayes estimator cannot be estimated consistently.
To fix this difficulty, Maiti, $\et$ (2014) suggested that $\si_i^2$ has an inverse gamma distribution.
It is interesting to point out that the resulting empirical Bayes (EB) estimator of $\xi_i$ shrinks both means and variances.
Since the EB includes integration with respect to $\si_i^2$, however, the EB cannot be expressed in closed forms. 
Thus one needs numerical integration to provide values of the EB.
Maiti, $\et$ (2014) derived a second-order unbiased estimator of the conditional mean squared error (cMSE) of the EB given $(y_i, V_i)$. 
However, one needs heavy numerical compuation to provide values of the estimator of cMSE.
For unconditional MSE of the EB, no computational algorithm was provided in Maiti, $\et$ (2014), because the computation may be much harder.

\medskip
In this paper, we consider to approximate the Bayes estimator in the Fay-Herriot random dispersion model given in Maiti, $\et$ (2014).
Aprroximating the joint probability density function, we suggest the  the approximated Bayes estimator 
$$
\xi_i^{AB}=\z_i^T\bbe + \Big(1-{1\over 1+ \tau^2 (n_i+1+\al)/(V_i+\ga)}\Big)(y_i-\z_i^T\bbe),
$$
where $\al$ and $\ga$ are model parameters in the distribution of $\si_i^2$.
Since
$$
{V_i+\ga \over n_i+1+\al } = {n_i +1\over n_i+1+\al} { V_i\over n_i+1} + \Big(1- {n_i +1\over n_i+1+\al}\Big) {\ga \over \al},
$$
the estimator $\xi_i^{AB}$ is a dual shrinkage estimator with shrinking $y_i$ towards $\z_i^T\bbe$ and shrinking $V_i/(n_i+1)$ towards $\ga/\al$.
This approximation is valid in the case of large $n_i$, but we want to use this closed-form estimator even for small $n_i$.
For the purpose, we need to evaluate the estimation error.
Since $\xi_i^{AB}$ includes the model parameters $\bbe$, $\tau^2$, $\al$ and $\ga$, we estimate $\bbe$ with the generalized least squares estimator based on the approximated pdf and the other parameters $\tau^2$, $\al$ and $\ga$ via the moment methods.
We show the consistency of the suggested estimators for the model parameters.
Plugging-in the consistent estimators in $\xi_i^{AB}$ yields the approximated empirical Bayes (AEB) estimator $\hxi^{AEB}_i$.
The uncertainty of the AEB is measured via the the unconditional mean squared errors (MSE), and we obtain a second-order unbiased estimator of the MSE via the single parametric bootstrap method.

\medskip
In this paper,  we also treat the benchmark problem.
A potential difficulty of the AEB estimators $\hxi^{AEB}_i$ for small areas is that the overall estimate for a larger geographical area, which is constructed by a (weighted) sum of $\hxi^{AEB}_i$, is not necessarily equal to the corresponding direct estimate like the overall sample mean.
For instance, we consider the weighted mean $\yo_w=\sum_{j=1}^m w_i y_j$ for nonnegative constants $w_j$'s satisfying $\sum_{j=1}^m w_j=1$.
Then, we want to find predictors $\de_i$'s which satisfy the benchmark constraint $\sum_{j=1}^m w_j \de_j = \yo_w$. 
A solution of the benchmark problem is the constrained Bayes estimation suggested by Ghosh (1992) and Datta, Ghosh, Steorts and Maples (2011).
Using this approach, we suggest the benchmarked predictor based on $\hxi^{AEB}_i$ given by
$$
\de_i^{CAB}= \hxi_i^{AEB} + {w_i \over \sum_{j=1}^mw_j^2}\Big\{ \yo_w - \sum_{j=1}^m w_j \hxi_j^{AEB}\Big\}.
$$
A second-order unbiased estimator of the MSE of this constrained approximate Bayes estimator is derived.

\medskip
The paper is organized as follows:
A setup of the Fay-Herriot random dispersion model and the approximated Bayes estimator are given in Section \ref{sec:main}.
The estimators of the model parameters are also given there.
In Section \ref{sec:MSE}, the approximated empirical Bayes (AEB) estimator is evaluated in terms of the MSE, and the second-order unbiased estimator is suggested.
The benchmark problem is discussed in Section \ref{sec:bench}.
In Section \ref{sec:sim}, we investigate the performance of the proposed procedures through simulation and empirical studies.
Concluding remarks are given in Section \ref{sec:remark} and the technical proofs are given in the Appendix.

\section{Area-level Model and Estimation of Model Parameters}
\label{sec:main}

\subsection{Fay-Herriot random dispersion model and an approximated predictor}

For $m$ small areas, let $(y_i, V_i/n_i)$ be the pair of mean estimate and variance estimate for the $i$-th small area, $i=1, \ldots, m$, where $n_i$ is degrees of freedom.
Suppose that there exist $p-1$ covariates which are denoted by $\z_i=(z_{i1}, \ldots, z_{ip})$ with $z_{i1}=1$. 
Then we consider the following heteroscedastic area-level model with random dispersions:
\begin{equation}
\begin{split}
y_i\mid \xi_i, \si_i^2 \sim& \Nc(\xi_i, \si_i^2),
\\
\xi_i \sim& \Nc(\z_i^T\bbe, \tau^2),
\\
V_i/\si_i^2 \mid \si_i^2 \sim& \chi_{n_i}^2,
\\
\si_i^{-2} \sim& Ga(\al/2, 2/\ga),
\end{split}
\label{eqn:model}
\end{equation}
where $(y_1, V_1),  \ldots, (y_m, V_m)$ are mutually independent.
We call it the Fay-Herriot Random Dispersion model (hereafter, FHRD model). 
Here, $Ga(\al/2, 2/\ga)$ denotes a gamma distribution with mean $\al/\ga$ and variance $2\al/\ga^2$.
The unknown parameters are denoted by $\bom=(\bbe^T, \tau^2, \al, \ga)^T$ for $\bbe=(\be_1, \ldots, \be_p)^T$.

\medskip
Let $\eta_i=1/\si_i^2$ and $C_i=1/[2\pi 2^{n_i/2}\Ga(n_i/2)]$.
The jonint pdf of $(y_i, V_i, \xi_i, \eta_i)$ is 
\begin{equation}
\begin{split}
f_i(y_i, V_i, \xi_i, \eta_i)=& C_i{(\ga/2)^{\al/2} \over \tau \Ga(\al/2)}
V_i^{n_i/2-1}\eta_i^{(n_i+1+\al)/2-1}
\\
&\times\exp\Big[ -{\eta_i\over 2}\{(y_i-\xi_i)^2+V_i+\ga\} - {1\over 2\tau^2}(\xi_i-\z_i^T\bbe)^2\Big].
\end{split}
\label{eqn:jpdf}
\end{equation}
It is noted that
\begin{align*}
\eta_i &(y_i-\xi_i)^2 + \tau^{-2}(\xi_i-\z_i^T\bbe)^2+\eta_i(V_i+\ga)
\\
&=(\eta_i+\tau^{-2})\{\xi_i- \xi_i^M(\eta_i) \}^2 + \eta_i\Big\{ V_i+\ga+{1\over \tau^2\eta_i+1}(y_i-\z_i^T\bbe)^2\Big\},
\end{align*}
where
\begin{equation}
\xi_i^M(\eta_i) = \z_i^T\bbe + \Big(1-{1\over \tau^2\eta_i+1}\Big)(y_i-\z_i^T\bbe).
\label{eqn:pmean}
\end{equation}
Then, the Bayes estimator of $\xi_i$ is described as
\begin{equation}
\xi_i^B = E[\xi_i\mid y_i, V_i] 
= \z_i^T\bbe + \Big(1- E\Big[{1\over \tau^2\eta_i+1}\mid y_i, V_i\Big] \Big)(y_i-\z_i^T\bbe),
\label{eqn:Bayes}
\end{equation}
where
\begin{equation}
E\Big[{1\over \tau^2\eta_i+1}\mid y_i, V_i\Big]
={\int_{0}^\infi (\tau^2\eta_i+1)^{-1} f_i(y_i, V_i, \eta_i)\dd \eta_i \over
\int_{0}^\infi  f_i(y_i, V_i, \eta_i)\dd \eta_i},
\label{eqn:BS}
\end{equation}
for the marginal pdf of $f_i(y_i, V_i, \eta_i)$ given by
\begin{equation}
\begin{split}
f_i(y_i, V_i, \eta_i)=& C_i{(\ga/2)^{\al/2} \over \Ga(\al/2)}
V_i^{n_i/2-1}\eta_i^{(n_i+1+\al)/2-1}
{\sqrt{2\pi}\over \sqrt{\tau^2\eta_i+1}}\\
&\times\exp\Big[ -{\eta_i\over 2}\Big\{ V_i+\ga + {1 \over \tau^2\eta_i+1}(y_i-\z_i^T\bbe)^2\Big\}\Big].
\end{split}
\label{eqn:jpdf1}
\end{equation}
When $\si_i^2=1/\eta_i$ is fixed and unknown, it may be estimated with $V_i/n_i$.
Then from (\ref{eqn:pmean}), one gets the estimator
$$
\z_i^T\bbe + \Big(1- {1\over \tau^2(n_i/V_i)+1} \Big)(y_i-\z_i^T\bbe),
$$
which is not very stable for small $n_i$ due to the estimation error in $V_i/n_i$.
The Bayes estimator (\ref{eqn:Bayes}) can fix this undesirable property.
However, we resort to numerical integration to obtain the Bayes estimator and the empirical Bayes estimator.
It may be computationally harder to evaluate the mean squared error of the empirical Bayes estimator.

\medskip
We want to suggest another estimator with a closed form.
To this end, we begin by integrating out the joint density (\ref{eqn:jpdf}) with respect to $\eta_i$.
Then the marginal pdf of $(y_i, V_i, \xi_i)$ is written as
\begin{align*}
h_i(y_i, V_i, \xi_i)=& C_i{(\ga/2)^{\al/2} \over \tau \Ga(\al/2)}
V_i^{n_i/2-1}\exp\Big[ -{1\over 2\tau^2}(\xi_i-\z_i^T\bbe)^2\Big]
\\
&\times \Ga\Big({n_i+1+\al\over 2}\Big) \Big({ 2\over (y_i-\xi_i)^2+V_i+\ga}\Big)^{(n_i+1+\al)/2}.
\end{align*}
Based on the density $h_i(y_i, V_i, \xi_i)$, the Bayes estimator of $\xi_i$ is also expressed as $\xi_i^B = E[\xi_i\mid y_i, V_i]$.
We here consider to approximate the marginal pdf $h_i(y_i, V_i, \xi_i)$.
It is noted that
\begin{align*}
\Big(&{ 2\over (y_i-\xi_i)^2+V_i+\ga}\Big)^{(n_i+1+\al)/2}
\\
&=\Big({ 2\over V_i+\ga}\Big)^{(n_i+1+\al)/2}
\exp\Big[ - {n_i+1+\al \over 2}\log\Big( 1 + {(y_i-\xi_i)^2\over V_i+\ga}\Big)\Big],
\end{align*}
Then, the function $\log\{ 1 + (y_i-\xi_i)^2/(V_i+\ga)\}$ is approximated as
\begin{equation}
\log\{ 1 + (y_i-\xi_i)^2/(V_i+\ga)\}\approx (y_i-\xi_i)^2/(V_i+\ga).
\label{eqn:approx}
\end{equation}
This approximation can be guaranteed when $n_i$ is large.
However, we use this approximation for small $n_i$ as well, and derive estimators of the unknown parameters and predictors for $\xi_i$ based on this approximation.

\medskip
Using this approximation, we can rewrite the pdf $h_i(y_i, V_i, \xi_i)$ as
\begin{align*}
h_i^*(y_i, V_i, \xi_i)=& C_i{(\ga/2)^{\al/2} \over \tau \Ga(\al/2)}
V_i^{n_i/2-1} \Big({2\over V_i+\ga}\Big)^{(n_i+1+\al)/2}\Ga\Big({n_i+1+\al\over 2}\Big)
\\
&\times  \exp\Big[ -{1\over 2\tau^2}(\xi_i-\z_i^T\bbe)^2 - {A_i\over 2} (y_i-\xi_i)^2 \Big],
\end{align*}
for $A_i =  (n_i+1+\al)/(V_i+\ga)$.
It is noted that $h_i^*(y_i, V_i, \xi_i)$ is not a pdff.
Since
$$
{1\over \tau^2}(\xi_i-\z_i^T\bbe)^2 + A_i (y_i-\xi_i)^2
= { 1+ \tau^{2}A_i \over \tau^2}\Big( \xi_i - {\z_i^T\bbe+\tau^2A_i y_i \over 1+\tau^2A_i}\Big)^2
+ {A_i\over 1+\tau^2 A_i}(y_i-\z_i^T\bbe)^2,
$$
we get the approximated Bayes estimator of $\xi_i$ given by
\begin{equation}
\xi^{AB}_i=\xi^{AB}_i(\bbe, \bth) = {\z_i^T\bbe+\tau^2A_i y_i \over 1+\tau^2A_i}
=\z_i^T\bbe + (1-B_i)(y_i-\z_i^T\bbe),
\label{eqn:ABayes}
\end{equation}
where $\bth=(\tau^2, \al, \ga)^T$ and 
$$
B_i = B_i(\bth, V_i)= {1\over 1+\tau^2A_i} = {1\over 1+ \tau^2 (n_i+1+\al)/(V_i+\ga)}.
$$
It is noted that this is not the Bayes estimator, but the approximated Bayes estimator when the approximation (\ref{eqn:approx}) is valid.
Since the approximated Bayes estimator has a simple and reasonable form, however, we shall use this estimator even if this approximation is not appropriate.
The following proposition implies that the approximated Bayes estimator $\xi_i^{AB}$ has less shrinkage than the Bayes estimator $\xi_i^B$ given in (\ref{eqn:Bayes}).
The proof is given in the Appendix.

\begin{prp}
\label{prp:1}
The shrinkage function $B_i$ in $\xi_i^{AB}$ is less than the shrinkage function $(\ref{eqn:BS})$ in the Bayes estimator $\xi_i^B$, namely,
$$
E\Big[{1\over \tau^2\eta_i+1}\mid y_i, V_i\Big]
\geq B_i = {1\over 1+ \tau^2 (n_i+1+\al)/(V_i+\ga)}.
$$
\end{prp}

\subsection{Estimation of the model parameters}

We now provide estimators of the model parameters $\bbe$, $\tau^2$, $\al$ and $\ga$.

\medskip
{\bf [1] Estimation of $\bbe$.}\ \ 
Integrating out $h_i^*(y_i, V_i, \xi_i)$ with respect to $\xi_i$, we have
\begin{align*}
h_i^*(y_i, V_i)=& C_i{(\ga/2)^{\al/2} \over \Ga(\al/2)}
V_i^{n_i/2-1} \Big({2\over V_i+\ga}\Big)^{(n_i+1+\al)/2}\Ga\Big({n_i+1+\al\over 2}\Big)
\\
&\times \sqrt{2\pi} \sqrt{B_i} \exp\Big[ -{1\over 2\tau^2}(1-B_i)(y_i-\z_i^T\bbe)^2 \Big].
\end{align*}
Let $\ell^*=\ell^*(\bbe, \tau^2,\al,\ga)=\sum_{i=1}^m \ell_i^*$ for $\ell_i^*=\log h_i^*(y_i, V_i)$.
Since 
$$
2{\partial \ell^*\over \partial \bbe}={1\over \tau^2}\sum_{i=1}^m(1-B_i)(y_i-\z_i^T\bbe)\z_i,
$$
we get the estimator
\begin{equation}
\bbet=\bbet(\bbe, \tau^2, \al, \ga)=\Big(\sum_{j=1}^m(1-B_j)\z_j\z_j^T\Big)^{-1}\sum_{j=1}^m (1-B_j)\z_j y_j,
\label{eqn:GLS}
\end{equation}
which is the generalized least squares (GLS) estimator of $\bbe$.

\medskip
{\bf [2] Estimation of $\tau^2$.}\ \ 
To estimate $\tau^2$, we consider the expectation $E[(y_i-\z_i^T\bbe)^2/(V_i+\ga)]$.
The conditional expectation of $(y_i-\z_i^T\bbe)^2$ given $V_i$ is decomposed as
$$
E[(y_i-\z_i^T\bbe)^2\mid V_i]
= E[(y_i-\xi_i)^2\mid V_i] + 2E[(y_i-\xi_i)(\xi_i-\z_i^T\bbe)\mid V_i] 
+ E[(\xi_i-\z_i^T\bbe)^2\mid V_i].
$$
Since $E[(y_i-\xi_i)^2\mid V_i, \eta_i, \xi_i]=1/\eta_i$, $E[(y_i-\xi_i)(\xi_i-\z_i^T\bbe)\mid V_i, \eta_i, \xi_i]=0$ and $E[(\xi_i-\z_i^T\bbe)^2\mid V_i, \eta_i]=\tau^2$, it is seen that
$$
E[(y_i-\z_i^T\bbe)^2\mid V_i]
=E[\eta_i^{-1}\mid V_i] + \tau^2.
$$
The joint pdf of $(V_i, \eta_i)$ is 
\begin{equation}
f_i(V_i,\eta_i) = {\ga^{\al/2}V_i^{n_i/2-1} \over \Ga(n_i/2)\Ga(\al/2)2^{(n_i+\al)/2}}
\eta_i^{(n_i+\al)/2-1}e^{-(\eta_i/2)(V_i+\ga)},
\label{eqn:pdfve}
\end{equation}
so that the marginal pdf of $V_i$ is
\begin{equation}
f_i(V_i) = {\Ga((n_i+\al)/2)\over \Ga(n_i/2)\Ga(\al/2)}{\ga^{\al/2} V_i^{n_i/2-1}\over (V_i+\ga)^{(n_i+\al)/2}},
\label{eqn:pdfv}
\end{equation}
and the conditional pdf of $\eta_i$ given $V_i$ is 
\begin{equation}
\eta_i \mid V_i \sim Ga\Big( {n_i+\al\over 2}, {2\over V_i+\ga}\Big).
\label{eqn:post}
\end{equation}
Thus, one gets
$$
E[\eta_i^{-1} \mid V_i] = {V_i+\ga \over n_i+\al-2},
$$
which implies that
\begin{equation}
E[(y_i-\z_i^T\bbe)^2 \mid V_i] = {V_i+\ga \over n_i+\al-2} +  \tau^2.
\label{eqn:ye}
\end{equation}
Thus, from this equality, we consider the moment $E[ (y_i-\z_i^T\bbe)^2/(V_i+\ga)]$, which is 
$$
E\Big[ {(y_i-\z_i^T\bbe)^2\over V_i+\ga}\Big] = {1\over n_i+\al-2} + E\Big[{ \tau^2\over V_i+\ga}\Big].
$$
To calculate the moments of $V_i$ from the marginal pdf (\ref{eqn:pdfv}), the following equality is useful:
In general, for real numbers $\ell$ and $k$, it can be shown that
\begin{equation}
E[{V_i^\ell\over (V_i+\ga)^k}] = {\Ga((n_i+\al)/2)\over \Ga((n_i+\al)/2+k)}
{\Ga(n_i/2+\ell)\over \Ga(n_i/2)}{\Ga(\al/2+k-\ell)\over \Ga(\al/2)}\ga^{\ell-k}.
\label{eqn:A1}
\end{equation}
For $\ell=0$ and $k=1$, we have $E[1/(V_i+\ga)]=\al/\{\ga(n_i+\al)\}$, so that
$$
E\Big[ {(y_i-\z_i^T\bbe)^2\over V_i+\ga}\Big] ={1\over n_i+\al-2} + { \al/\ga\over n_i+\al}\tau^2.
$$
When $\al$ and $\ga$ are known and $\bbe$ is estimated by the ordinary least squares (OLS) estimator $\bbeh_{OLS}=(\sum_{j=1}^m\z_j\z_j^T)^{-1}\sum_{j=1}^m \z_j y_j$, 
this gives us the estimator
\begin{equation}
\tah^2 = \Bigl(\sum_{i=1}^m{\al/\ga\over n_i+\al}\Bigr)^{-1}
\sum_{i=1}^m \Big\{ {(y_i-\z_i^T\bbeh_{OLS})^2\over V_i+\ga}- {1\over n_i+\al-2}\Big\}.
\label{eqn:tauh}
\end{equation}

\smallskip
{\bf [3] Estimation of $\al$.}\ \ 
Concerning the estimation of $\al$, we concentrate on the marginal pdf (\ref{eqn:pdfv}) of $V_i$.
Since $\log f_i(V_i)$ is expressed as
$$
\log f_i(V_i) = \log\Ga({n_i+\al\over 2}) - \log\Ga({n_i\over 2})-\log\Ga({\al\over 2})
+ {\al\over 2}\log \ga + {n_i -2 \over 2}\log V_i - {n_i+\al\over 2} \log(V_i+\ga),
$$
we have 
$$
2 {\partial \over \partial \al}\log f_i(V_i) = \psi({n_i+\al\over 2}) - \psi({\al\over 2})
+ \log \ga - \log(V_i+\ga),
$$
where $\psi(\cdot)$ is the digamma function given by $\psi(x)=\Ga'(x)/\Ga(x)$.
Since $E[ \partial f_i(V_i)/\partial\al]=0$, one gets
\begin{equation}
E[\log(V_i+\ga)] = \psi({n_i+\al\over 2}) - \psi({\al\over 2}) + \log \ga.
\label{eqn:al1}
\end{equation}

We here note the following equality.
For real numbers $\ell$ and $k$, it can be shown that
\begin{align}
E\Big[{V_i^\ell\over (V_i+\ga)^k}\log(V_i+\ga)\Big] =& {\Ga((n_i+\al)/2)\over \Ga((n_i+\al)/2+k)}
{\Ga(n_i/2+\ell)\over \Ga(n_i/2)}{\Ga(\al/2+k-\ell)\over \Ga(\al/2)}\ga^{\ell-k}
\non\\
&\times
\Bigl\{ \psi({n_i+\al\over 2}+k) - \psi({\al\over 2}+k-\ell)+\log\ga\Bigr\}.
\label{eqn:A2}
\end{align}
For $\ell=1$ and $k=1$, we have
$$
E\Big[{V_i\over V_i+\ga}\log(V_i+\ga)\Big] 
= {n_i \over n_i +\al} \Bigl\{ \psi({n_i+\al\over 2}+1) - \psi({\al\over 2}) + \log \ga\Bigr\}.
$$
Since the digamma function has the property that $\psi(x+1)=\psi(x)+1/x$, it follows from (\ref{eqn:al1}) that
\begin{equation}
E\Big[{V_i\over V_i+\ga}\log(V_i+\ga)\Big] 
= {n_i \over n_i +\al} \Bigl\{ E[\log(V_i+\ga)] + {2\over n_i+\al}\Bigr\}.
\label{eqn:al2}
\end{equation}
This can be rewritten as
\begin{align*}
\al^2 E\Big[{V_i\over V_i+\ga}\log(V_i+\ga)\Big] &+ \al E\Big[n_i {V_i-\ga\over V_i+\ga}\log(V_i+\ga)\Big] 
\\
&- n_i^2E\Big[{\ga\over V_i+\ga}\log(V_i+\ga)\Big] - 2n_i=0,
\end{align*}
which yields an estimator of $\al$.
In fact, we can suggest the estimator as the solution of the quadratic equation
\begin{align}
\al^2 \sum_{i=1}^m{V_i\over V_i+\ga}\log(V_i+\ga) &+ \al \sum_{i=1}^m n_i {V_i-\ga\over V_i+\ga}\log(V_i+\ga) 
\non\\
&- \sum_{i=1}^m n_i \Big\{ {n_i\ga\over V_i+\ga}\log(V_i+\ga) + 2\Big\}=0.
\label{eqn:al}
\end{align}

\smallskip
{\bf [4] Estimation of $\ga$.}\ \ 
Concerning the estimation of $\ga$, from (\ref{eqn:A1}), it follows that
$$
E\Big[{V_i\over V_i+\ga}\Big] 
= {n_i \over n_i +\al}.
$$
Thus, one gets the estimator of $\ga$ as the solution of the equation
\begin{equation}
\sum_{i=1}^m {V_i \over V_i+\ga}=\sum_{i=1}^m {n_i\over n_i+\al}.
\label{eqn:ga}
\end{equation}

\section{Evaluation of Uncertainty of Prediction}
\label{sec:MSE}

Substituting the estimators of $\bbe$, $\tau^2$, $\al$ and $\ga$ into (\ref{eqn:ABayes}), we get the predictor
\begin{equation}
\hxi^{AEB}_i=\xi_i^{AB}(\bbeh, \tah^2, \alh, \gah) = \z_i^T\bbeh + (1-\Bh_i)(y_i-\z_i^T\bbeh),
\label{eqn:pred}
\end{equation}
where
\begin{align}
\Bh_i =& B_i(\tah^2, \alh, \gah) = {1\over 1+ \tah^2 (n_i+1+\alh)/(V_i+\gah)}.
\label{eqn:Bh}
\\
\bbeh=&\Big(\sum_{j=1}^m(1-\Bh_j)\z_j\z_j^T\Big)^{-1}\sum_{j=1}^m (1-\Bh_j)\z_j y_j,
\label{eqn:EGLS}
\end{align}
We call it the approximated empirical Bayes estimator.
It is noted that the term $(V_i + \gah)/(n_i+1+\alh)$ in $\Bh_i$ is expressed as
$$
{V_i + \gah \over n_i+1+\alh} = {n_i +1\over n_i+1+\alh}{V_i\over n_i+1} + \Bigl(1-{n_i+1\over n_i+1+\alh}\Bigr){\gah\over \alh},
$$
which shrinks $V_i/(n_i+1)$ towards the target $\gah/\alh$.
Thus, the predictor $\hxi^{AEB}_i$ is a double shrinkage procedure such that $y_i$ and $V_i/(n_i+1)$ are shrunken towards $\z_i^T$ and $\gah/\alh$, respectively.
In this seciton, we derive a second-order unbiased estimator of the mean squared error (MSE) of $\hxi^{AEB}_i$.

\medskip
We begin by rewriting the predictor as
$$
\hxi^{AEB}_i = \{(1-B_i)y_i+B_i\z_i^T\bbe\}
- \{ (\Bh_i-B_i)(y_i-\z_i^T\bbe)-\Bh_i\z_i^T(\bbeh-\bbe)\}.
$$
Thus, the MSE of $\hxi^{AEB}_i$ is decomposed as
\begin{equation}
MSE(\hxi^{AEB}_i) = E[ (\xih_i-\xi_i)^2]= g_1 + g_2 -2g_3,
\label{eqn:MSE}
\end{equation}
where
\begin{equation}
\begin{split}
g_1=& E[ \{(1-B_i)y_i+B_i\z_i^T\bbe - \xi_i \}^2],
\\
g_2=&E[\{ (\Bh_i-B_i)(y_i-\z_i^T\bbe)-\Bh_i\z_i^T(\bbeh-\bbe)\}^2],
\\
g_3 =& E[\{(1-B_i)y_i+B_i\z_i^T\bbe - \xi_i \} \{ (\Bh_i-B_i)(y_i-\z_i^T\bbe)-\Bh_i\z_i^T(\bbeh-\bbe)\}].
\end{split}
\label{eqn:g}
\end{equation}

To evaluate $g_1$, $g_2$ and $g_3$, we use the following theorem under the assumption (A):
For notational simplicity, let $\bth=(\tau^2, \al, \ga)^T$ and $\bthh=(\tah^2, \alh, \gah)^T$.
Also, let $\bom=(\bbe^T, \bth^T)^T$ and $\bomh=(\bbeh^T, \bthh^T)^T$.

\medskip
\noindent
{\bf Assumption (A)}

\medskip
(A1) There exist $\underline{n}$ and $\overline{n}$ such that $\underline{n}\leq n_i \leq \overline{n}$ for $i=1, \ldots, m$.
The dimension $p$ is bounded.

\medskip
(A2) The matrix $m^{-1}\sum_{i=1}^m \z_i\z_i^T$ converges to a positive definite matrix.

\begin{thm}
\label{thm:moment}
Assume the condition {\rm (A)} and $n_i+\al>4$ for $i=1, \ldots, m$.
Then, $E[\bomh-\bom]=O(m^{-1})$ and $E[(\bomh-\bom)(\bomh-\bom)^T]=O(m^{-1})$.
Also, the conditional moments given $y_i, V_i$ satisfy that $E[\bomh-\bom\mid y_i, V_i]=E[\bomh-\bom]+o_p(m^{-1})$ and $E[(\bomh-\bom)(\bomh-\bom)^T \mid y_i, V_i]=E[(\bomh-\bom)(\bomh-\bom)^T]+o_p(m^{-1})$.
\end{thm}

We first evaluate $g_1$. 
Since $(1-B_i)y_i+B_i\z_i^T\bbe - \xi_i =(1-B_i)(y_i-\z_i^T\bbe) - (\xi_i-\z_i^T\bbe)$ and $B_i$ is a function of $V_i$, it is seen that 
\begin{align*}
g_1=& E[(1-B_i)^2(y_i-\z_i^T\bbe)^2] + E[(\xi_i-\z_i^T\bbe)^2] - 2 E[(1-B_i)(y_i-\z_i^T\bbe)(\xi_i-\z_i^T\bbe)]
\\
=&
E[ (1-B_i)^2(\si_i^2 + \tau^2) + \tau^2 - 2 (1-B_i)\tau^2].
\end{align*}
Note that $E[\si_i^2\mid V_i]=E[1/\eta_i\mid V_i]=(V_i+\ga)/(n_i-2+\al)$.
Thus, one gets $g_1 =E[G(\bth, V_i)]$, where
\begin{equation}
G(\bth, V_i)={V_i+\ga \over n_i-2+\al}(1-B_i)^2 + \tau^2B_i^2,
\label{eqn:G}
\end{equation}
which is of order $O_p(1)$.
We here rewrite $g_1$ as
$$
g_1=E[G(\bthh, V_i)] - E[G(\bthh, V_i) - G(\bth, V_i)] = g_{11} - g_{12}. \quad {\rm (say)}
$$
An exact unbiased estimator of $g_{11}$ is $G(\bthh, V_i)$.
Concerning $g_{12}$, the Taylor series expansion give us the approximation as
\begin{align*}
g_{12}=& E\Bigl[ (\bthh-\bth)^T{\partial \over \partial \bth} G(\bth, V_i)\Bigr] + 
{1\over 2} E\Bigl[  (\bthh-\bth)^T\Bigl( {\partial^2 \over \partial \bth\partial \bth^T} G(\bth, V_i)\Bigr) (\bthh-\bth)\Bigr] + O(m^{-3/2}).
\end{align*}
Since $\bomh-\bom = O_p(m^{-1/2})$ from Theorem \ref{thm:moment}, it is clear that the second term in $g_{12}$ is of order $O(m^{-1})$.
For the first term, it follows from Theorem \ref{thm:moment} that
\begin{align*}
E\Bigl[ (\bthh-\bth)^T{\partial \over \partial \bth} G(\bth, V_i) \mid y_i, V_i \Bigr]
=&
E[ (\bthh-\bth)^T\mid y_i, V_i] {\partial \over \partial \bth} G(\bth, V_i)
\\
=&
E[(\bthh-\bth)^T] {\partial \over \partial \bth} G(\bth, V_i) + o_p(m^{-1}),
\end{align*}
which is of order $O_p(m^{-1})$.
This shows that $g_{12}=O(m^{-1})$.

\medskip
For $g_2$, it is clear that $g_2=O(m^{-1})$.
For $g_3$, it is noted that
\begin{align*}
g_3 =&E[\{(1-B_i)y_i+B_i\z_i^T\bbe - E[\xi_i\mid y_i, V_i] \} \{ (\Bh_i-B_i)(y_i-\z_I^T\bbe)- B_i\z_i^T(\bbeh-\bbe)\}]
\\
&- E[\{(1-B_i)y_i+B_i\z_i^T\bbe - E[\xi_i\mid y_i, V_i] \} (\Bh_i-B_i) \z_i^T(\bbeh-\bbe)\}].
\end{align*}
The second term is of order $O(m^{-1})$.
For the first term, 
$$
E[\{(1-B_i)y_i+B_i\z_i^T\bbe - E[\xi_i\mid y_i, V_i] \} \{ E[\Bh_i-B_i\mid y_i, V_i] (y_i-\z_i^T\bbe)- B_i\z_i^T E[\bbeh-\bbe\mid y_i, V_i]\}],
$$
which is approximated as
$$
E[\{(1-B_i)y_i+B_i\z_i^T\bbe - E[\xi_i\mid y_i, V_i] \} \{ E[\Bh_i-B_i] (y_i-\z_i^T\bbe)- B_i\z_i^T E[\bbeh-\bbe]\}] + o(m^{-1}).
$$
This shows that $g_3=O(m^{-1})$.
Hence, we get the following proposition.

\begin{prp}
\label{prp:MSE}
Assume the condition {\rm (A)} and $n_i+\al>4$ for $i=1, \ldots, m$.
Then, the MSE of the predictor $\hxi^{AEB}_i$ is decomposed as
$$
MSE(\hxi^{AEB}_i) = g_{11} + \{- g_{12} + g_2 -2g_3\},
$$
where $g_{11}=O(1)$, $g_{12}=O(m^{-1})$, $g_2=O(m^{-1})$ and $g_3=O(m^{-1})$.
\end{prp}

We next estimate the MSE of $\hxi^{AEB}_i$.
An exact unbiased estimator of $g_{11}$ is given by
$$
\gh_{11} = G(\bthh, V_i)={V_i+\gah \over n_i-2+\alh}(1-\Bh_i)^2 + \tah^2\Bh_i^2.
$$
To provide second-order unbiased estimators of $g_{12}$, $g_2$ and $g_3$, we use the parametric bootstrap method.
Let $(y_i^*, V_i^*)$, $i=1, \ldots, m$, be a bootstrap sample generated from the model:
\begin{equation}
\begin{split}
y_i^*\mid \xi_i^*, \eta_i^* \sim& \Nc(\xi_i^*, 1/\eta_i^* ),
\\
\xi_i^* \sim& \Nc(\z_i^T\bbeh, \tah^2),
\\
V_i^* \eta_i^* \mid \eta_i^* \sim& \chi_{n_i}^2,
\\
\eta_i^* \sim& Ga(\alh/2, 2/\gah),
\end{split}
\label{eqn:bmodel}
\end{equation}
where $\bbeh$, $\tah^2$, $\alh$ and $\gah$ are estimators constructed from the original model (\ref{eqn:model}).
The bootstrap estimators $\bbeh^*$, $\tah^*$, $\alh^*$ and $\gah^*$ are calculated via the same manner as in $\bbeh$, $\tah^2$, $\alh$ and $\gah$ except that the bootstrap estimators are calculated based on $(y_i^*, V_i^*)$'s instead of $(y_i, V_i)$'s.
Then we can estimate $g_{12}$, $g_{2}$ and $g_3$ with
\begin{equation}
\begin{split}
g_{12}^* =&E^*[G(\bthh^*, V_i^*) - G(\bthh, V_i^*)],
\\
g_2^*=&E^*[\{ (\Bh_i^*-B_i^*)(y_i^*-\z_i^T\bbeh)-\Bh_i^*\z_i^T(\bbeh^*-\bbeh)\}^2],
\\
g_3^* =& E^*[\{(1-B_i^*)y_i^*+B_i^*\z_i^T\bbeh - \xi_i^* \} \{ (\Bh_i^*-B_i^*)(y_i^*-\z_i^T\bbeh)-\Bh_i^*\z_i^T(\bbeh^*-\bbeh)\}],
\end{split}
\label{eqn:bg}
\end{equation}
where $\bthh^*=(\tah^*,\alh^*,\gah^*)^T$, $B_i^*=B_i(\bthh,V_i^*)$ and $\Bh_i^*=B_i(\bthh^*, V_i^*)$.
It can be seen that these are second-order unbiased estimators.

\begin{prp}
\label{prp:mse}
Assume the condition {\rm (A)} and $n_i+\al>4$ for $i=1, \ldots, m$.
Then, a second-order unbiased estimator of the MSE of $\hxi^{AEB}_i$ is
\begin{equation}
mse(\hxi^{AEB}_i) = \gh_{11} + \{- g_{12}^* + g_2^* -2g_3^*\},
\label{eqn:mse_AEB}
\end{equation}
where $E[\gh_{11}]=g_{11}$, $E[g_{12}^*]=g_{12}+o(m^{-1})$, $E[g_2^*]=g_2+o(m^{-1})$ and $E[g_3^*]=g_3+o(m^{-1})$.
\end{prp}

\section{Benchmarked Prediction}
\label{sec:bench}

In this section, we consider the benchmark problem which imposes a constraint on predictors for small areas.
The benchmarked predictors are derived and an approxomated unbiased estimator of their MSE is provided.

\medskip
Although the predictors $\xih_i$ in (\ref{eqn:pred}) give reliable estimates for $\xi_i$ by borrowing strength from the surrounding areas, we are faced with a potential difficulty of the predictor.
That is, the overall estimate for a larger geographical area, which is constructed by a (weighted) sum of $\xih_i$, is not necessarily equal to the corresponding direct estimate like the overall sample mean.
To describe it specifically, let $w_j$'s be nonnegative constants such that $\sum_{j=1}^m w_j=1$.
Suppose that the mean of the total areas is estimated by the weighted sum of $y_j$'s, $\sum_{j=1}^m w_j y_j$.
Then, the benchmark problem is described as an issue of finding estimators $\de_j$ such that
\begin{equation}
\sum_{j=1}^m w_j \de_j = \sum_{j=1}^m w_i y_j \equiv \yo_w.
\label{eqn:bench}
\end{equation}
A solution of the benchmark problem is the constrained Bayes estimation suggested by Ghosh (1992) and Datta, Ghosh, Steorts and Maples (2011), who considered the minimization of $\sum_{i=1}^m E[(\de_i-\xi_i)^2\mid Data]$ under the constraint (\ref{eqn:bench}).
Using the Lagrange multiplier method, one gets the constrained Bayes estimator
$$
\de_i^{CB}= E[\xi_i\mid y_i, V_i] + {w_i \over \sum_{j=1}^mw_j^2}\Big\{ \yo_w - \sum_{j=1}^m w_j E[\xi_j\mid y_j, V_j]\Big\}.
$$
Since the Bayes estimator $E[\xi_i\mid y_i, V_i]$ cannot be expressed in a closed form, we replace it with the approximated empirical Bayes estiamtor $\xih_i$ given in (\ref{eqn:pred}).
The resulting benchmarked predictor is
\begin{equation}
\de_i^{CAB}= \hxi_i^{AEB} + {w_i \over \sum_{j=1}^mw_j^2}\Big\{ \yo_w - \sum_{j=1}^m w_j \hxi_j^{AEB}\Big\},
\label{eqn:Bpred}
\end{equation}
which is here called the constrained approximate Bayes estimator.

\medskip
To evaluate the uncertainty of $\de_i^{CAB}$, we derive a second-order unbiased estimator of thh MSE.
The MSE of $\de_i^{CAB}$ is decomposed as
\begin{align*}
E[(\de_i^{CAB}-\xi_i)^2] =& E[(\hxi_i^{AEB}-\xi_i)^2]
\\
&+ {w_i^2 \over (\sum_{j=1}^mw_j^2)^2}E\Big[(\yo_w-\sum_{j=1}^mw_j\hxi_j^{AEB})^2\Big]
+2 {w_i \over \sum_{j=1}^m w_j^2} J,
\end{align*}
where
$$
J= E\Big[(\hxi_i^{AEB}-\xi_i)\Big(\yo_w-\sum_{j=1}^mw_j\hxi_j^{AEB}\Big)\Big].
$$
The second-order unbiased estimator of the first term $E[(\hxi_i^{AEB}-\xi_i)^2]$ is given in Proposition \ref{prp:mse}.
Clearly, an exact unbiased estimator of $E[(\yo_w-\sum_{j=1}^mw_j\hxi_j^{AEB})^2]$ is $(\yo_w-\sum_{j=1}^mw_j\hxi_j^{AEB})^2$.
In Theorem \ref{thm:Bmse} given below, we verify that $J=O(1)$.
Then, we can estimate $J$ based on the bootstrap sample given in (\ref{eqn:bmodel}) as
$$
J^* = E^*\Big[(\hxi_i^{AEB*}-\xi_i^*)\Big(\yo_w^*-\sum_{j=1}^mw_j\hxi_j^{AEB*}\Big)\Big],
$$
which satisfies that $E[J^*]=J + o(1)$.

\begin{thm}
\label{thm:Bmse}
Assume the condition {\rm (A)} and $n_i+\al>4$ for $i=1, \ldots, m$.
Also assume that $\sum_{j=1}^m w_j^2/m$ converges to a non-zero constant.
Then, $J=O(1)$ and a second-order unbiased estimator of the MSE of $\de_i^{CAB}$ is
\begin{equation}
mse(\de_i^{CAB}) = mse(\hxi_i^{AEB}) + {w_i^2 \over (\sum_{j=1}^mw_j^2)^2}\Big[(\yo_w-\sum_{j=1}^mw_j\hxi_j^{AEB})^2\Big]
+2 {w_i \over \sum_{j=1}^m w_j^2} J^*,
\label{eqn:mse_CAB}
\end{equation}
where $mse(\hxi_i^{AEB})$ is given in $(\ref{eqn:mse_AEB})$.
That is, $E[mse(\de_i^{CAB})]=E[(\de_i^{CAB}-\xi_i)^2]+o(m^{-1})$.
\end{thm}

The proof of Theorem \ref{thm:Bmse} is given in the Appendix.

\section{Numerical and Empirical Studies}
\label{sec:sim}
In this section, we investigate performances of the procedures suggested in the previous sections through numerical and empirical studies.

\subsection{Simulation study}

Here we investigate finite sample performances of the estimators in the Fay-Herriot random dispersion (FHRD) model and the second-order unbiased estimators for the unconditional MSEs by the Monte Carlo simulation. 
Comparing the performances of the approximated Bayes estimator $\xi_i^{AB}$ given in (\ref{eqn:ABayes}) with those of the Bayes estimator $\xi_i^B$ given in (\ref{eqn:Bayes}), we check goodness of the approximation we applied.

\medskip
We conduct simulation experiments as we specified true model, so simulation data is generated by FHRD model (\ref{eqn:model}). 
Throughout the simulations, the true value of $\bbe$ and $\ga$ are $\be = 10$ and $\ga = 1$.
For each of $m,\tau^{2},\al$, we examined two cases; $m = 30$ or $60$, $\tau^{2} = 1$ or $4$ and $\al = 1$ or $4$. 
For simplicity, we set $\z_{i} = 1$, $p=1$ and $n_{i}=10$ for all areas and cases. 
Thus, there are eight cases of similations for the variety of $m$, $\al$ and $\tau^{2}$.

\medskip
We forst compute numerical values of MSE of the Bayes and approximate Bayes estimators $\xi_{i}^{B}$ and $\xi_{i}^{AB}$ with
$$
{\rm MSE_{i}}=\frac1K\sum_{k=1}^K\left(\hat{\xi}_{i}^{(k)}- \xi_{i}^{(k)}\right)^2
$$
for $K=5,000$ where $\xih_i^{(k)}$ and $\xi_i^{(k)}$ are the estimator and the true value of $\xi_i$ in the $k$-th replication for $k=1,\ldots,K$. 
To investigate the loss which arises from approximation (\ref{eqn:approx}), we compare the approximated Bayes estimator $\xi_i^{AB}$ with the Bayes estimator $\xi_i^B$ in terms of biases and true MSEs under known model parameters. The results of the simulation are reported in Table \ref{table:1}.

\begin{table}[!htb]
\caption{Biases and square roots of MSE of $\xi_{i}^{AB}$ and $\xi_{i}^{B}$ in the FHRD model under known model parameters}\label{table:1}
\begin{center}
\begin{tabular}{ccccccc}
\toprule
Estimator & $\al,\tau^{2}$&   &    $Bias$ (m=30)  &  $SRMSE$ (m=30) &    $Bias$ (m=60)  &  $SRMSE$ (m=60)\\
\midrule
    & $1, \ 1$&&  0.002 & 0.818 & -0.002 & 0.822\\
$\xi_{i}^{B}$ & $1, \ 4$&& 0.003 & 1.347 & -0.005 & 1.346\\
    & $4, \ 1$&& -0.005 & 0.528 & 0.000 & 0.528\\
     & $4, \ 4$&& -0.005 & 0.629 & -0.002 & 0.628\\
\midrule
    & $1, \ 1$&&  0.002 & 0.824 & -0.002 & 0.828\\
$\xi_{i}^{AB}$ & $1, \ 4$&& 0.003 & 1.355 & -0.005 & 1.355\\
    & $4, \ 1$&& -0.006 & 0.530 & 0.001 & 0.530\\
     & $4, \ 4$&& -0.005 & 0.631 & -0.002 & 0.629\\
\bottomrule
\end{tabular}
\end{center}
\end{table}

It is observed from Table \ref{table:1} that the difference between $\xi_{i}^{B}$ and $\xi_{i}^{AB}$ is tiny. 
Even though $\xi_{i}^{AB}$ is dominated by $\xi_{i}^{B}$ as expected, the difference of the biases is little except the case $(\al,\tau^{2}) = (4,1)$. 
Moreover, the largest difference between the two SRMSEs is $0.009$ for $m = 60$ and $(\al,\tau^{2}) = (1,4)$. 
Thus, the approximation little affects the bias and MSE. 

\medskip
We next investigate finite sample performances of the estimators for the model parameters. 
In particular, it is worth remarking the estimation of $\al$.
Initially, we used the maximum likelihood estimator of $\al$ from the marginal likelihood as Maiti et al. (2014).
The MLE can be obtained by solving the equation based on the digamma functions.
However, the numerical solutions for the MLE yield large variability.
To avoid such an instable performance of the MLE, in this paper, we suggest the new consistent estimator given in (\ref{eqn:al}).
The performances of the suggested estimators for $\be$, $\tau^2$, $\al$ and $\ga$ are reported in Table \ref{table:2}, where  means and standard diviations via simulation with $1,000$ replications are given.
Table \ref{table:2} shows that the estimator of $\be$ is almost unbiased and has small standard deviation. 
For other estimators, both biases and standard deviations are moderated as $m$ increases. 
Espetially, the suggested estimator (\ref{eqn:al}) of $\al$ provides stable estimates and a good performance.

\begin{table}[!htb]
\caption{Results of the estimation for model parameters $\be$, $\tau^{2}$, $\al$ and $\ga$ in the FHRD model for $\be=10$ and $\ga=1$(Standard deviations are shown in the parentheses)}\label{table:2}
\begin{center}
\begin{tabular}{ccccccc}
\toprule
$m$ & $\al,\tau^{2}$&   &    $\hat{\be}$  &  $\hat{\tau^{2}}$ &    $\hat{\al}$  &  $\hat{\ga}$\\
\midrule
    & $1, \ 1$&&  10.001(0.331) & 0.912(0.658) & 1.041(0.160) & 1.092(0.262)\\
$m = 30$ & $1, \ 4$&& 9.997(0.495) & 3.658(1.834) & 1.038(0.157) & 1.092(0.269)\\
    & $4, \ 1$&& 10.000(0.217) & 0.950(0.361) & 4.067(0.538) & 1.013(0.085)\\
     & $4, \ 4$&& 10.006(0.384) & 3.876(1.235) & 4.063(0.543) & 1.015(0.085)\\
\midrule
    & $1, \ 1$&&  9.999(0.226) & 0.944(0.483) & 1.135(0.189) & 1.203(0.346)\\
$m = 60$ & $1, \ 4$&& 10.006(0.345) & 3.883(1.321) & 1.018(0.114) & 1.036(0.167)\\
    & $4, \ 1$&& 10.000(0.148) & 0.977(0.254) & 4.029(0.377) & 1.006(0.061)\\
     & $4, \ 4$&& 9.998(0.268) & 3.928(0.880) & 4.036(0.383) & 1.005(0.062)\\
\bottomrule
\end{tabular}
\end{center}
\end{table}

\medskip
Finally, we compare the second-order unbiased estimator of the MSE of $\xih_{i}^{AEB}$ with the true MSE. 
Concerning the MSE, the true value is calculated via simulation with $R=5,000$ replications as \ref{eqn:MSE}. 
Then, the mean values of the estimator for the MSE and their Percentage Relative Bias (RB) are calculated based on $T=1,000$ simulation runs with each $1,000$ bootstrap samples, where RB is defined by
\begin{align*}
{\rm RB}_i&=100\frac{T^{-1}\sum_{t=1}^T\widehat{\rm MSE}_i^{(t)}-{\rm MSE}_i}{{\rm MSE}_i},
\end{align*}
for the MSE estimate $\widehat{\rm MSE}^{(t)}$ in the $t$-th replication for $t=1,\ldots,T$. Note that both true MSE and its estimator are calculated based on our estimates of the model parameters.
Table \ref{table:3} reports values of the true MSE, the second-order unbiased estimator $\widehat{MSE}$ and the corresponding Percentage Relative Biases. 
It is observed that the MSE estimates $\widehat{\rm MSE}$ are close to the true values of the MSE, and their relative bias are small for both $m=30$ and $60$. 
Thus, the second-order unbiased estimator of the MSE performs well as an estimator of the MSE of $\xih_i^{AEB}$ in the FHRD model.

\begin{table}[!htb]
\caption{Relative biases of the MSE estimator for $\xih_I^{AEB}$ in the FHRD model}
\label{table:3}
\begin{center}
\begin{tabular}{ccccccc}
\toprule
Size & $\al,\tau^{2}$&   &    $MSE$  &  $\widehat{MSE}$ &     RB(\%) \\
\midrule
    & $1, \ 1$&&  0.996 & 0.952 & -4.379\\
$m = 30$ & $1, \ 4$&&  2.22 & 2.243 & 1.042\\
    & $4, \ 1$&&  0.312 & 0.323 & 3.580\\
     & $4, \ 4$&&  0.416 & 0.414 & -0.443\\
\midrule
    & $1, \ 1$&&  0.835 & 0.833 & -0.233\\
$m = 60$ & $1, \ 4$&&  2.026 & 2.011 & -0.719\\
    & $4, \ 1$&&  0.294 & 0.298 & 1.284\\
     & $4, \ 4$&&  0.399 & 0.405 & 1.424\\
\bottomrule
\end{tabular}
\end{center}
\end{table}

\subsection{Illustrative examples}

We apply the approximated empirical Bayes estimator and the estimator of the MSE to the data in the Survey of Family Income and Expenditure (SFIE) in Japan.

\medskip
In this study, we use the data of the spending items ^^ Education' and ^^ Health' in the survey in 2014.
For the spending item ^^ Education', the annual average spending (scaled by 1,000 Yen) at each capital city of 47 prefectures in Japan is denoted by $y_i$ for $i=1, \ldots, 47$, and each variance $V_i$ is calculated based on data of the spending ^^ Education' at the same city in the past consecutive eight years.
Although the annual average spendings in SFIE are reported every year, the sample sizes are around 50 for most prefectures. 
We apply the same manner to the spending item ^^ Helth' to create $y_i$ and $V_i$.
The data of the item ^^ Education' have high variability, but those of the item ^^ Health' have relatively lower one.

\medskip
In addition to the SFIE data, we can use data in the National Survey of Family Income and Expenditure (NSFIE) for 47 prefectures. 
Since NSFIE is based on much larger sample than SFIE, the annual average spendings in NSEDI are more reliable, but this survey has been implemented every five years. 
In this study, we use the data of the spending items ^^ Education' and ^^ Health' of NSFIE in 2009 as covariates $z_i$ for $i=1, \ldots, 47$.
Thus, we apply the FHRD model (\ref{eqn:model}) to these examples, where $\z_i^T\bbe=\be_0+z_i \be_1$.

\medskip
We calculated the predicted values of $\xih_i^{AEB}$ and $\xih_i^{EB}$ and the estimates of the MSE of $\xih_i^{AEB}$ with the estimates $V_i$. 
These values in seven prefectures around Tokyo are reported in Tables \ref{table:4} and \ref{table:5} for the ^^ Education' and ^^ Health' data. 
For the ^^ Education' data, the estimates of the model parameters are $\beh_{0} = 15.711$, $\beh_{1} = 0.140$, $\tah^{2} = 12.069$, $\alh = 2.050$ and $\gah = 2.764$.
For the ^^ Health' data, the estimates of the model parameters are $\beh_{0} = 8.819$, $\beh_{1} = 0.192$, $\tah^{2} = 5.497$, $\alh = 9.502$ and $\gah = 2.109$.
As seen from the tables, the ^^ Education' data have more variability than the ^^ Health' data. 
The approximated empirical Bayes estimator $\xih_i^{AEB}$ returns almost similar values as the empirical Bayes estimator $\xih_i^{EB}$ does. 
Both estimators do not shrink $y_i$ so much.
It is also from Table \ref{table:4} that $\xih_i^{AEB}$ and $\xih_i^{EB}$ shrink $y_i$ more toward $\beh_0+z_i\beh_1$ when the values of $V_i$ are larger. 
The MSE estimates of $\xih_i^{AEB}$ give large values for large $V_i$'s.


\begin{table}[!htb]
\caption{Predicted values and the MSE estimates for the ^^ Education' data}\label{table:4}
\begin{center}
\begin{tabular}{ccccccccccc}
\toprule
Prefecture && $V_{i}$ & $y_{i}$&$\z_i^T\bbeh$ &  $\xih_{i}^{AEB}$ & $\xih_{i}^{EB}$  & $\widehat{MSE}_{AEB}$ \\
\midrule
Ibaraki   &&  4.210 & 21.972 & 17.873&  21.768 & 21.851 & 1.098\\
Tochigi  &&  4.974 & 21.883 & 18.102&  21.675 & 21.768 & 1.157\\
Gunma    &&  11.157 & 14.115 & 17.933&   14.475 &  14.287 & 1.772\\
Saitama &&  72.622 & 32.608 & 19.309&  27.805 & 29.064 & 5.868\\
Chiba    &&  26.419 & 21.554 & 18.751&  21.050 & 21.274 & 3.144\\
Tokyo   &&  13.091 & 22.037 & 19.337&  21.750 & 21.915 & 2.084\\
Kanagawa &&  16.266 & 22.321 & 18.494&  21.843 & 22.106 & 2.260\\
\bottomrule
\end{tabular}
\end{center}
\end{table}

\begin{table}[!htb]
\caption{Predicted values and the MSE estimates for the ^^ Health' data}\label{table:5}
\begin{center}
\begin{tabular}{ccccccccccc}
\toprule
Prefecture && $V_{i}$ & $y_{i}$&$\z_i^T\bbeh$ & $\xih_{i}^{AEB}$ & $\xih_{i}^{EB}$  & $\widehat{MSE}_{AEB}$ \\
\midrule
Ibaraki   &&  1.160 & 10.351 & 10.946&  10.369 & 10.410 & 0.211\\
Tochigi  &&  3.964 & 11.759 & 11.080&  11.720 & 11.572 & 0.369\\
Gunma    &&  3.444 & 8.737 & 10.307&  8.818 &  9.139 & 0.349\\
Saitama &&  0.920 & 11.133 & 11.316& 11.138 & 11.146 & 0.194\\
Chiba    &&  3.720 & 12.808 & 11.150& 12.718 & 12.388 & 0.358\\
Tokyo   &&  1.161 & 13.803 & 10.959& 13.714 & 13.477 & 0.208\\
Kanagawa &&  0.479 & 14.496 & 11.088& 14.411 & 14.293 & 0.163\\
\bottomrule
\end{tabular}
\end{center}
\end{table}

\section{Concluding Remarks}
\label{sec:remark}
In the Fay-Herriot random dispersion (FHRD) model, we have derived the approximated empirical Bayes (AEB) estimator with the closed form and provided the second-order unbiased estimator of the MSE of the AEB estimator via the single parametric bootstrap method.
Through various simulation experiments and empirical studies, it has been shown that the difference between the AEB estimator and the empirical Bayes estimator given in Maiti, $\et$ (2014) is small.
This means that the AEB estimator is useful irrespective of validity of the approximation.
It has been observed that the estimators suggested in this paper for the model parameters have good performances.
Especiall, our estimator of $\al$ is described in the closed form, and it performs well.

\normalsize
\medskip
\bigskip
\noindent
{\bf Acknowledgments.}\ \

Research of the second author was supported in part by Grant-in-Aid for Scientific Research  (23243039 and 26330036) from Japan Society for the Promotion of Science.

\bigskip
\appendix
\section{Appendix}

We here give proofs of Proposition \ref{prp:1} and Theorems \ref{thm:moment} and \ref{thm:Bmse}.

\medskip
{\bf [1] Proof of Proposition \ref{prp:1}.}\ \ 
It follows from (\ref{eqn:BS}) and (\ref{eqn:jpdf1}) that
$$
E\Big[{1\over \tau^2\eta_i+1}\mid y_i, V_i\Big]
={\int_{0}^\infi (\tau^2\eta_i+1)^{-1} D_i(\eta_i) g_i^*(\eta_i) \dd \eta_i \over
\int_{0}^\infi D_i(\eta_i) g_i^*(\eta_i) \dd \eta_i},
$$
where 
\begin{align*}
g_i^*(\eta_i)=& \eta_i^{(n_i+1+\al)/2-1} \exp[-\eta_i(V_i+\ga)/2],
\\
D_i(\eta_i) =& {1\over \sqrt{\tau^2\eta_i+1}}\exp\Big[ - {\eta_i \over 2(\tau^2\eta_i+1)}(y_i-\z_i^T\bbe)^2\Big].
\end{align*}
It is noted that $D_i(\eta_i)$ is a decreasing function of $\eta_i$.
Then, we first show that
\begin{equation}
{\int_{0}^\infi (\tau^2\eta_i+1)^{-1} D_i(\eta_i) g_i^*(\eta_i) \dd \eta_i \over
\int_{0}^\infi D_i(\eta_i)g_i^*(\eta_i) \dd \eta_i}
\geq
{\int_{0}^\infi (\tau^2\eta_i+1)^{-1} g_i^*(\eta_i) \dd \eta_i \over
\int_{0}^\infi g_i^*(\eta_i) \dd \eta_i}.
\label{eqn:pprp1}
\end{equation}
This inequaity is equivalent to
\begin{equation}
E_*[ (\tau^2\eta_i+1)^{-1} D_i(\eta_i)] \geq E_*[ (\tau^2\eta_i+1)^{-1} ] E_*[D_i(\eta_i)],
\label{eqn:pprp2}
\end{equation}
where $E_*[\cdot]$ is the expectation with respect to the pdf $C_i g_i^*(\eta_i)$ for some constant $C_i$.
The inequality (\ref{eqn:pprp2}) is equivalent to
\begin{align*}
Cov_*&((\tau^2\eta_i+1)^{-1}, D_i(\eta_i))
\\
&=E_*\Big[ \big\{ (\tau^2\eta_i+1)^{-1} - E_*[ (\tau^2\eta_i+1)^{-1}]\big\} \big\{ D_i(\eta_i) - E_*[D_i(\eta_i)]\big\}\Big] \geq 0,
\end{align*}
which holds since both $(\tau^2\eta_i+1)^{-1}$ and $D_i(\eta_i)$ are decreasing in $\eta_i$.
Thus, one gets the ineqaulity (\ref{eqn:pprp1}).

Since $(\tau^2\eta_i+1)^{-1}$ is a convex function of $\eta_i$, the Jensen inequality is applied to show the inequality
$$
{\int_{0}^\infi (\tau^2\eta_i+1)^{-1} g_i^*(\eta_i) \dd \eta_i \over
\int_{0}^\infi g_i^*(\eta_i) \dd \eta_i}
\geq
\Big(\tau^2 {\int_{0}^\infi \eta_i g_i^*(\eta_i) \dd \eta_i \over
\int_{0}^\infi g_i^*(\eta_i) \dd \eta_i} + 1\Big)^{-1}.
$$
Noting that $g_i^*(\eta_i)$ is proportional to the pdf of $Ga((n_i+1+\al)/2, 2/(V_i+\ga))$, we can see that ${\int_{0}^\infi \eta_i g_i^*(\eta_i) \dd \eta_i / \int_{0}^\infi g_i^*(\eta_i) \dd \eta_i} = (n_i+\al+1)/(V_i+\ga)$.
Namely,
\begin{equation}
{\int_{0}^\infi (\tau^2\eta_i+1)^{-1} g_i^*(\eta_i) \dd \eta_i \over
\int_{0}^\infi g_i^*(\eta_i) \dd \eta_i}
\geq
{ 1 \over \tau^2 (n_i+\al+1)/(V_i+\ga) + 1}.
\label{eqn:pprp3}
\end{equation}
Combining (\ref{eqn:pprp1}) and (\ref{eqn:pprp3}) proves Proposition \ref{prp:1}.

\bigskip
{\bf [2] Proof of Theorem \ref{thm:moment}.} \ \ 
For notational simplicity, let $\bth=(\th_1, \th_2,\th_3)^T=(\tau^2, \al, \ga)^T$ and $\bthh=(\thh_1, \thh_2,\thh_3)^T=(\tah^2, \alh, \gah)^T$.
Since $\bbeh$ is expressed as $\bbeh=\bbet(\bthh)$, the Taylor series expansion gives that
\begin{align}
\bbeh - \bbe =& \bbet - \bbe +
\sum_{a=1}^3 \Bigl\{ \Bigl(\sum_{j=1}^m (1-B_j)\z_j\z_j^T\Big)^{-1}\Big(\sum_{j=1}^m {\partial B_j \over \partial \th_a}\z_j\z_j^T\Bigr) \bbet 
\label{eqn:bep}
\\
&\qquad\qquad\qquad - \Bigl(\sum_{j=1}^m (1-B_j)\z_j\z_j^T\Big)^{-1} \sum_{j=1}^m {\partial B_j \over \partial \th_a}\z_j\z_j^{T} \bbe^* \Big\} (\thh_a-\th_a) +o_p(m^{-1}),
\non
\end{align}
where $\bbe^*=\{\sum_{j=1}^m (\partial B_j/\partial \th_a)\z_j\z_j^T\}^{-1} \sum_{j=1}^m (\partial B_j/\partial \th_a)\z_j y_j$.
Since $E[(y_i-\z_i^T\bbe)^2 \mid V_i] = (V_i+\ga)/( n_i+\al-2) + \tau^2$ from (\ref{eqn:ye}), it is observed that 
\begin{align*}
E[(\bbet-\bbe)(\bbet-\bbe)^T\mid V_i]=&
E\Big[\Big(\sum_{j=1}^m (1-B_j)\z_j\z_j^T\Big)^{-1} \Big(\sum_{j=1}^m \z_j\z_j^T(1-B_j)^2\Big\{ {V_j+\ga\over n_j+\al-2}+\tau^2\Big\}\Big)
\\
&\qquad\times \Big(\sum_{j=1}^m (1-B_j)\z_j\z_j^T\Big)^{-1}\mid V_i\Big],
\end{align*}
which implies that $E[(\bbet-\bbe)(\bbet-\bbe)^T]=O(m^{-1})$, and $\bbet-\bbe=O_p(m^{-1/2})$ under the condition (A) for $n_j+\al>2$.
Also, it is seen that
\begin{align}
E[&(\bbet-\bbe)(\bbet-\bbe)^T\mid y_i, V_i]=
E[(\bbet-\bbe)(\bbet-\bbe)^T] 
\label{eqn:bep1}
\\
&+ E\Big[\Big(\sum_{j=1}^m (1-B_j)\z_j\z_j^T\Big)^{-1} \z_i\z_i^T(1-B_i)^2 \Big( (y_i-\z_i^T\bbe)^2- \Big\{ {V_j+\ga\over n_j+\al-2}+\tau^2\Big\}\Big) 
\non\\
&\qquad\times \Big(\sum_{j=1}^m (1-B_j)\z_j\z_j^T\Big)^{-1} \mid y_i, V_i\Big],\non
\end{align}
which shows that $E[(\bbet-\bbe)(\bbet-\bbe)^T\mid y_i, V_i]=E[(\bbet-\bbe)(\bbet-\bbe)^T] +O_p(m^{-2})$ and $\bbet-\bbe\mid y_i, V_i =O_p(m^{-1/2})$.
Clearly, $E[\bbet-\bbe]=\zero$, and it is observed that
\begin{align*}
E[\bbet - \bbe\mid y_i, V_i]=&
E\Big[\Big\{\sum_{j=1}^m (1-B_j)\z_j\z_j^T\Big\}^{-1} \sum_{j=1}^m (1-B_j)\z_j (y_j-\z_j^T\bbe) \mid y_i, V_i\Big]
\\
=&
E\Big[\Big\{\sum_{j=1}^m (1-B_j)\z_j\z_j^T\Big\}^{-1} \mid y_i, V_i\Big] (1-B_i)\z_i (y_i-\z_i^T\bbe),
\end{align*}
which means that $E[\bbet - \bbe\mid y_i, V_i]=O_p(m^{-1})$.
Similarly, we can show that $E[\bbe^* - \bbe\mid y_i, V_i]=O_p(m^{-1})$ and $\bbe^* - \bbe\mid y_i, V_i=O_p(m^{-1/2})$.
Thus, from (\ref{eqn:bep}), we can verify that $E[(\bbeh-\bbe)(\bbeh-\bbe)^T]=O(m^{-1})$,  $E[\bbeh-\bbe\mid y_i, V_i]=+o_p(m^{-1})$ and $E[(\bbeh-\bbe)(\bbeh-\bbe)^T \mid y_i, V_i]=E[(\bbeh-\bbe)(\bbeh-\bbe)^T]+o_p(m^{-1})$, provided $\bthh=(\thh_1, \thh_2, \thh_3)^T$ satisfy the results given in Theorem \ref{thm:moment}.

\medskip
We now prove the results concerning $\bthh$ in Theorem \ref{thm:moment}.
Let $\F(\bth)=(F_1(\bth), F_2(\bth), F_3(\bth))^T$, where 
\begin{align*}
F_1(\bth)=& {1\over m}\sum_{j=1}^m\Big\{ {\al/\ga\over n_j+\al}\tau^2 + {1\over n_j+\al-2}-{(y_j-\z_j^T\bbeh_{OLS})^2\over V_j+\ga}\Big\},
\\
F_2(\bth)=& {1\over m}\sum_{j=1}^m\Big\{ \al^2{V_j\over V_j+\ga}\log(V_j+\ga) + \al n_j{V_j-\ga\over V_j+\ga}\log(V_j+\ga) - {n_j^2\ga\over V_j+\ga}\log(V_j+\ga) - 2n_j\Big\},
\\
F_3(\bth)=& {1\over m}\sum_{j=1}^m\Big\{ {V_j\over V_j+\ga} - {n_j\over n_j+\al}\Big\}.
\end{align*}
Since $\F(\bthh)=\zero$, the consistency of $\bthh$ follows from the Cramer method explained in Jiang (2010).
It is noted that for $a=1, 2, 3$, 
$$
F_a(\bthh)=F_a(\bth) + \Big({\partial F_a(\bth)\over \partial \bth^T}\Big) (\bthh-\bth)
+{1\over 2}(\bthh-\bth)^T{\partial^2 F_a \over \partial\bth\partial\bth^T}(\bthh-\bth) + o_p(\Vert\bthh-\bth\Vert^2),
$$
which yields
\begin{align}
\bthh-\bth =& - \Big({\partial \F(\bth)\over \partial \bth^T}\Big)^{-1}\F(\bth)
- {1\over 2} \Big({\partial \F(\bth)\over \partial \bth^T}\Big)^{-1}\Col_a\Big( (\bthh-\bth)^T{\partial^2 F_a \over \partial\bth\partial\bth^T}(\bthh-\bth) \Big)
\non\\
&+o_p(\Vert\bthh-\bth\Vert^2),
\label{eqn:thp}
\end{align}
where $\Col_a(x_a)=(x_1, x_2, x_3)^T$ and $\Vert\bthh-\bth\Vert^2=(\bthh-\bth)^T(\bthh-\bth)$.
Hence, it is sufficient to show that $E[\{F_a(\bth)\}^2]=O(m^{-1})$, $E[\{F_a(\bth)\}^2\mid y_i, V_i]=O_p(m^{-1})$, $E[F_a(\bth)]=O_p(m^{-1})$ and $E[F_a(\bth)\mid y_i, V_i]=O_p(m^{-1})$ for each $a$, and $\partial \F(\bth)/ \partial \bth^T$ converges to a positive definite matrix.

\medskip
Concerning $F_1(\bth)$, note that 
$$
F_1(\bth)=F_1^*(\bbe, \bth) + {2\over m}\sum_{j=1}^m {y_j-\z_j^T\bbe \over V_j+\ga}\z_j^T(\bbeh_{OLS}-\bbe) + O_p(m^{-1}),
$$
where
$$
F_1^*(\bbe, \bth)= {1\over m}\sum_{j=1}^m\Big\{ {\al/\ga\over n_j+\al}\tau^2 + {1\over n_j+\al-2}-{(y_j-\z_j^T\bbe)^2\over V_j+\ga}\Big\}.
$$
It can be verified that $E[\{F_1(\bth)\}^2]=O(m^{-1})$, $E[\{F_1(\bth)\}^2\mid y_i, V_i]=O_p(m^{-1})$, $E[F_1(\bth)]=O_p(m^{-1})$ and $E[F_1(\bth)\mid y_i, V_i]=O_p(m^{-1})$ if there exists $E[\{(y_j-\z_j^T\bbe)^4/(V_j+\ga)^2]$.
Similarly, the corresponding properties of the moments for $F_2(\bth)$ and $F_3(\bth)$ can be demonstrated if there exists $E[\{\log(V_j+\ga)\}^2]$.
Thus, we need to check these moments.
It is noted that
\begin{align*}
E[\{(y_j-\z_j^T\bbe)^4\mid V_j] =&
E[(y_j-\xi_j)^4+6(y_j-\xi_j)^2(\xi_j-\z_j^T\bbe)^2 + (\xi_j-\z_j^T)^4\mid V_j]
\\
=& E\Big[ {3\over \eta_j^2} + {6\tau^2\over  \eta_j} + 3\tau^4\mid V_j\Big]
\\
=&
{3(V_j+\ga)^2\over (n_j+\al-2)(n_j+\al-4)} + {6\tau^2(V_j+\ga)\over n_j+\al-2} + 3\tau^4,
\end{align*}
so that $E[\{(y_j-\z_j^T\bbe)^4/(V_j+\ga)^2]$ is finite if $n_j+\al > 4$.
To investigate the existence of $E[\{\log(V_j+\ga)\}^2]$, we calculate both sides of $4E[-(\partial^2/\partial\al^2)\log f(V_j)]=4E[\{ (\partial/\partial\al)\log f(V_j)\}^2]$.
The RHS is equal to $E[\{\psi((n_j+\al)/2)-\psi(\al/2)+\log\ga - \log(V_j+\ga)\}^2]=\{\psi((n_j+\al)/2)-\psi(\al/2)+\log\ga\}^2 - E[\{\log(V_j+\ga)\}^2]$.
On the other hand, the LHS is $-\psi'((n_j+\al)/2)+\psi'(\al/2)$.
Hence,
$$
E[\{\log(V_j+\ga)\}^2] = \{\psi((n_j+\al)/2)-\psi(\al/2)+\log\ga\}^2 + \psi'((n_j+\al)/2)-\psi'(\al/2),
$$
which is finite for $\al>0$.

\medskip
Finally, we show that every entry of the matrix $\partial \F(\bth)/ \partial \bth^T$ converges  in probabiltiy.
For notational simplicity, let $F_{a(i)}(\bth)=\partial F_a(\bth)/\partial \th_i$ for $a=1, 2, 3$ and $i=1,2,3$.
Then, for $F_{1(i)}$, we have $F_{1(1)}=m^{-1}\sum_j \al/\{\ga(n_j+\al)\}$, and
$$
F_{1(2)}={1\over m}\sum_j\Big\{{n_j\tau^2/\ga\over (n_j+\al)^2}-{1\over (n_j+\al-2)^2}\Big\}, 
F_{1(3)}={1\over m}\sum_j\Big\{-{\tau^2/\ga^2\over n_j+\al}+{(y_j-\z_j^T\bbeh_{OLS})^2\over (V_j+\ga)^2}\Big\}.
$$
For $F_{2(i)}$, we have $F_{2(1)}=0$, and
\begin{align*}
F_{2(2)}=& {1\over m}\sum_j{\log(V_j+\ga)\over V_j+\ga}\{ 2\al V_j + n_j(V_j-\ga)\},
\\
F_{2(3)}=& - {1\over m}\sum_j{V_j\log(V_j+\ga)\over (V_j+\ga)^2}(\al+n_j)^2
+  {1\over m}\sum_j{\al+n_j\over (V_j+\ga)^2}\{ \al V_j - n_j\ga).
\end{align*}
For $F_{3(i)}$, we have $F_{3(1)}=0$, and 
$$
F_{3(2)}={1\over m}\sum_j {n_j \over (n_j+\al)^2}, \quad 
F_{3(3)}=- {1\over m}\sum_j {V_j \over (V_j+\ga)^2}.
$$
Thus, it can be seen that these converge to the limiting values of their expectations if there exist the moments $E[\{(y_j-\z_j^T\bbe)^4/(V_j+\ga)^2]$ and $E[\{\log(V_j+\ga)\}^2]$.
 are finite.
Therefore, the proof of Theorem \ref{thm:moment} is complete.

\medskip
{\bf [3] Proof of Theorem \ref{thm:Bmse}.} \ \ 
It is noted from (\ref{eqn:Bayes}) that $\xi_i^B = E[\xi_i\mid y_i, V_i] = \z_i^T\bbe + (1- E_i )(y_i-\z_i^T\bbe)$ for $E_i =E[(\tau^2\eta_i+1)^{-1}\mid y_i, V_i]$.
Then, $J$ can be rewritten as
\begin{align*}
J=& E\Big[(\hxi_i^{AEB}-\xi_i^B)\Big(\yo_w-\sum_{j=1}^mw_j\hxi_j^{AEB}\Big)\Big]
\\
=& E\Big[ (B_i-E_i)(y_i-\z_i^T\bbe)\sum_{j=1}^m w_jB_j(y_j-\z_j^T\bbe)\Big]
\\
&+E\Big[\Big\{\Bh_i\z_i^T(\bbeh-\bbe)-(\Bh_i-B_i)(y_i-\z_i^T\bbe)\Big\}\sum_{j=1}^mw_j\Bh_j(y_j-\z_j^T\bbeh)\Big]
\\
&+E\Big[(B_i-E_i)(y_i-\z_i^T\bbe)\sum_{j=1}^mw_j \Big\{(\Bh_j-B_j)(y_j-\z_j^T\bbeh)-B_j\z_j^T(\bbeh-\bbe)\Big\}\Big]
\\
=& J_1 + J_2 + J_3.\quad {\rm (say)}
\end{align*}
Noting that $E_i$ is a function of $V_i$ and $(y_i-\z_i^T\bbe)^2$, we can see that
$$
E[(B_i-E_i)(y_i-\z_i^T\bbe)(y_j-\z_j^T\bbe) \mid V_i, V_j] = 
\left\{\begin{array}{ll}
(B_i-E_i)(y_i-\z_i^T\bbe)^2 & {\rm for}\ i=j,
\\
0 & {\rm for}\ i\not= j.
\end{array}
\right.
$$
Thus, $J_1$ is evaluated as
\begin{align*}
J_1=& E\Big[w_iB_i(B_i-E_i)(y_i-\z_i^T\bbe)^2 + (B_i-E_i)\sum_{j\not= i}w_jB_j(y_i-\z_i^T\bbe)(y_j-\z_j^T\bbe)\Big]
\\
=& E[ w_iB_i(B_i-E_i)(y_i-\z_i^T\bbe)^2],
\end{align*}
which is of order $O(1)$.
Noting that $\bbeh-\bbe=O_p(m^{-1/2})$ and $\Bh_i-B_i=O_p(m^{-1/2})$, we can demonstrate that 
\begin{align*}
J_2=&E\Big[\Big\{B_i\z_i^T(\bbeh-\bbe)-(\Bh_i-B_i)(y_i-\z_i^T\bbe)\Big\}\sum_{j=1}^mw_jB_j(y_j-\z_j^T\bbe)\Big] + o(1)
\\
J_3=&E\Big[(B_i-E_i)(y_i-\z_i^T\bbe)\sum_{j=1}^mw_j \Big\{(\Bh_j-B_j)(y_j-\z_j^T\bbe)-B_j\z_j^T(\bbeh-\bbe)\Big\}\Big]+o(1).
\end{align*}
It is easily seen that 
$$
E\Big[\Big\{\sum_{j=1}^mw_jB_j(y_j-\z_j^T\bbe)\Big\}^2\mid V_1, \ldots, V_m\Big]
=E\Big[\sum_{j=1}^m w_j^2B_j^2 (y_j-\z_j^T\bbe)^2\mid V_1, \ldots, V_m\Big],
$$
which implies that $\sum_{j=1}^mw_jB_j(y_j-\z_j^T\bbe)=O_p(m^{-1/2})$.
Thus, one gets that $J_2=O(1)$.

Concernig $J_3$, it is noted that $\tah^2$ given in (\ref{eqn:tauh}) is approximated as 
$$
\tah^2 = \Bigl(\sum_{i=1}^m{\al/\ga\over n_i+\al}\Bigr)^{-1}
\sum_{i=1}^m \Big\{ {(y_i-\z_i^T\bbe)^2\over V_i+\ga}- {1\over n_i+\al-2}\Big\}
+O_p(m^{-1/2}),
$$
so that we can regard $\Bh_i$ as a function of $(y_i-\z_i^T\bbe)^2$ and $V_i$, $i=1, \ldots, m$.
Hence, 
$$
E\Big[(B_i-E_i)(y_i-\z_i^T\bbe)\sum_{j=1}^mw_j (\Bh_j-B_j)(y_j-\z_j^T\bbe)\Big]
=E[(B_i-E_i)(y_i-\z_i^T\bbe)^2w_i (\Bh_i-B_i)],
$$
which is of $O(m^{-1/2})$.
Since $\bbeh-\bbet=O_p(m^{-1/2})$, it is seen that
\begin{align*}
E\Big[&(B_i-E_i)(y_i-\z_i^T\bbe)\sum_{j=1}^mw_j B_j\z_j^T(\bbet-\bbe)\mid V_1, \ldots, V_m\Big]
\\
=&
E\Big[(B_i-E_i)(y_i-\z_i^T\bbe)\sum_{j=1}^mw_j B_j\z_j^T\Big(\sum_{a=1}^m(1-B_a)\z_a\z_a^T\Big)^{-1}
\\
&\times \sum_{a=1}^m(1-B_a)\z_a(y_a-\z_a^T\bbe)\mid V_1, \ldots, V_m\Big]
\\
=&
E\Big[(B_i-E_i)(y_i-\z_i^T\bbe)^2\sum_{j=1}^mw_j B_j\z_j^T\Big(\sum_{a=1}^m(1-B_a)\z_a\z_a^T\Big)^{-1}(1-B_i)\z_i\mid V_1, \ldots, V_m\Big],
\end{align*}
which is of order $O_p(1)$.
Hence, it is concluded that $J_3=O(1)$.
Therefore, the proof of Theorem \ref{thm:Bmse} is complete.

\end{document}

%% file: anot.tex
\def\hxi{\widehat{\xi}}

\def\al{{\alpha}}
\def\be{{\beta}}
\def\ga{{\gamma}}
\def\de{{\delta}}

\def\si{{\sigma}}

\def\th{{\theta}}

\def\bbe{{\text{\boldmath $\beta$}}}

\def\bom{{\text{\boldmath $\omega$}}}
\def\bth{{\text{\boldmath $\theta$}}}

\def\alh{{\widehat \al}}
\def\beh{{\widehat \be}}

\def\gah{{\hat \ga}}
\def\xih{{\hat \xi}}
\def\thh{{\hat \th}}

\def\tah{{\hat \tau}}

\def\bbeh{{\widehat \bbe}}

\def\bomh{{\widehat \bom}}
\def\bthh{{\widehat \bth}}

\def\bbet{{\widetilde \bbe}}

\def\Ga{{\Gamma}}

\def\z{{\text{\boldmath $z$}}}

\def\F{{\text{\boldmath $F$}}}

\def\yo{{\overline y}}

\def\gh{{\hat g}}

\def\Bh{{\widehat B}}

\def\Nc{{\cal N}}

\def\infi{{\infty}}

\def\Col{{\bf Col\,}}
\def\[{{\text{\boldmath $[$}}}
\def\]{{\text{\boldmath $]$}}}
\def\et{{\it et\, al.}}
\def\zero{{\bf\text{\boldmath $0$}}}

\def\dd{{\rm d}}

\def\|{{\,|\,}}

\def\/{{\Bigr/\!\!}}

\def\1r{{\rm (1)}}
\def\2r{{\rm (2)}}
\def\3r{{\rm (3)}}
\def\4r{{\rm (4)}}
\def\5r{{\rm (5)}}

\def\non{{\nonumber}}
